\documentclass[conference]{IEEEtran}
\IEEEoverridecommandlockouts
\usepackage{amsthm}
\usepackage{amsmath}
\usepackage{amssymb}
\usepackage{graphicx}
\usepackage{epstopdf}
\usepackage{algorithm}
\usepackage{algorithmic,color}
\usepackage{bm}

\usepackage{aas_macros}

\usepackage{framed}
\usepackage{bbold}
\usepackage{tikz}
\usetikzlibrary{positioning}

\def\argmin{\mathop{\rm argmin}}

\newcommand{\vect}[1]{\boldsymbol{#1}}

\usepackage{booktabs}
\usepackage{multirow}
\usepackage{pifont}

\newcommand{\tabL}{1.8mm}

\algsetup{indent=2em}

\begin{document}

\title{Quantifying Uncertainty in High Dimensional Inverse Problems by Convex Optimisation
\thanks{This work is supported by the UK Engineering and Physical Sciences Research Council (EPSRC) by grant EP/M011089/1, 
Science and Technology Facilities Council (STFC) ST/M00113X/1, and Leverhulme Trust.}
}


\author{\IEEEauthorblockN{1\textsuperscript{st} Xiaohao Cai}
\IEEEauthorblockA{\textit{Mullard Space Science Laboratory} \\
\textit{University College London (UCL)}\\
Surrey RH5 6NT, United Kingdom \\
x.cai@ucl.ac.uk}
\and
\IEEEauthorblockN{2\textsuperscript{nd} Marcelo Pereyra}
\IEEEauthorblockA{\textit{Maxwell Institute for Mathematical Sciences} \\
\textit{Heriot-Watt University}\\
Edinburgh EH14 4AS, United Kingdom \\
m.pereyra@hw.ac.uk}
\and
\IEEEauthorblockN{3\textsuperscript{rd} Jason D. McEwen}
\IEEEauthorblockA{\textit{Mullard Space Science Laboratory } \\
\textit{University College London (UCL)}\\
Surrey RH5 6NT, United Kingdom \\
jason.mcewen@ucl.ac.uk}
}

\maketitle
\begin{abstract} 
Inverse problems play a key role in modern image/signal processing methods.
However, since they are generally ill-conditioned or ill-posed due to lack of 
observations, their solutions may have significant intrinsic uncertainty.
Analysing and quantifying this uncertainty is very challenging, particularly in
high-dimensional problems and problems with non-smooth objective functionals (e.g. sparsity-promoting priors).
In this article, a series of strategies to visualise this uncertainty 
are presented, e.g. highest posterior density credible regions, and local credible intervals (\textit{cf.} error bars) for individual pixels and superpixels.
Our methods support non-smooth priors for inverse problems  and can be scaled to high-dimensional settings.
Moreover, we present strategies to automatically set regularisation parameters so that
the proposed uncertainty quantification (UQ) strategies become much easier to use.
Also, different kinds of dictionaries (complete and over-complete) are used to represent the image/signal and 
their performance in the proposed UQ methodology is investigated.
\end{abstract}

\section{Introduction} \label{sec:intro}
Inverse problems, like reconstruction (e.g. \cite{PMdCOW16,mce11}), denosing (e.g. \cite{ROF92}) or 
deblurring (e.g. \cite{SST10}), are important 
subjects in image/signal processing. Briefly speaking, the task is to recover 
images/signals that are as close to natural ones as possible from corrupted observations, 
e.g. noisy, blurry and/or incomplete data. The degraded quality of the observations
makes the associated inverse problems ill-conditioned or ill-posed, and therefore 
their solutions may have significantly intrinsic uncertainty (see e.g. \cite{CPM17,M15}).
Quantifying this kind of uncertainty, particularly for high-dimensional problems, is very challenging. 
This is the main focus in this article.

Briefly speaking, inverse problems in image/signal processing can be generally formulated as regularised estimation problems 
involving a data fidelity term and a regularisation term, which are often substantiated by using a statistical likelihood function 
and a prior distribution, respectively (see Section \ref{sec:problem} for more detail). 
Statistical sampling approaches like Markov Chain Monte Carlo (MCMC) sampling can in principle
recover the full posterior probability distribution of the image/signal from the given inverse problems, from which uncertainties (like error bars)
can then be quantified, see e.g. \cite{CPM17,M15,DMP16,R01}. However, due to the long computation time required to sample the full posterior distribution, 
these kind of methods can be extremely slow when data sets are large.
As an alternative, {\it maximum a posteriori} (MAP) estimation has been adopted as a standard approach, combined with convex optimisation 
techniques to compute the estimator in practice \cite{CP16,CP10}. Such approaches have led to significant improvements in estimation accuracy and computation time
in high dimensional scenarios.
Recently, the authors proposed a series of uncertainty quantification (UQ) strategies based on MAP estimation \cite{M16} 
to visualise uncertainties \cite{CPM17, CPM17b}, illustrated for radio interferometric (RI) imaging.
Note that the UQ techniques presented in \cite{M16,CPM17b}, which possess the advantages of MAP estimation as mentioned above,
are very different to the methods using MCMC sampling \cite{CPM17,M15,DMP16} 
but accurately approximate the same forms of UQ (see \cite{CPM17,M15,DMP16,M16, CPM17b} for more details).

In this article, based on the high-posterior-density (HPD) region approximation method \cite{M16}, 
we develop strategies to efficiently perform UQ in general image/signal processing inverse problems.
In particular, general dictionaries are considered in prior knowledge, and the involved regularisation parameter, which controls the strength of the prior knowledge 
in inverse problems and plays a key role in UQ analyses, is estimated automatically and used for the subsequent UQ analyses.
Simulations are implemented to demonstrate the validity and effectiveness of the
UQ strategies in terms of computing local credible intervals (\textit{cf.} error bars or Bayesian confidence intervals) for individual pixels and superpixels.

\section{Problem Formulation}\label{sec:problem}
Let $\vect y \in \mathbb{C}^M$ represent the observed data set (an image/signal degraded and/or transformed by
some operators [e.g. Fourier transform, blur  operator], noise,  and/or information loss), which satisfies
${\vect y}=\bm{\mathsf{\Phi}} {\vect x} + {\vect n}$,
where $\bm{\mathsf{\Phi}} \in \mathbb{C}^{M\times N}$, ${\vect x} \in \mathbb{R}^N$, and ${\vect n} \in \mathbb{C}^M$ 
denote, respectively, a problem related operator, the clean image/signal, and noise.
Without loss of generality, we subsequently consider independent and identically distributed (i.i.d.) Gaussian noise. 
Under a basis or dictionary (e.g., a wavelet basis or an over-complete frame \cite{CMW12})
$\bm{\mathsf{\bm{\mathsf{\Psi}}}} \in \mathbb{C}^{N\times L}$, $\vect{x}$ can be represented by  
${\vect x} = \bm{\mathsf{\Psi}} {\vect a} = \sum_{i} \bm{\mathsf{\Psi}}_i a_i$,
where vector ${\vect a} = (a_1, \cdots, a_L)^\top$ represents the synthesis coefficients
of ${\vect x}$ under $\bm{\mathsf{\Psi}}$. In particular, ${\vect x}$ is said to be compressible if many coefficients of $\vect{a}$ are small; 
natural images are generally compressible for appropriate choices of $\bm{\mathsf{\Psi}}$ \cite{D06}.

In practice, $\vect y$ is only observed partially or with limited resolution
and thus solving 
\begin{equation}\label{eqn:y}
{\vect y}=\bm{\mathsf{\Phi}} \vect x + {\vect n} \quad {\rm or} \quad 
	{\vect y}=\bm{\mathsf{\Phi}} \bm{\mathsf{\Psi}} {\vect a} + {\vect n}
\end{equation}
for $\vect x$ presents an ill-posed inverse problem. In particular, when $\bm{\mathsf{\Phi}}$ is an identity operator,
blur operator (e.g., Gaussian), or transformation (e.g., Fourier or Radon), the above inverse problem goes to 
image/signal denosing, deblurring, or reconstruction, respectively (see e.g. \cite{T96,CPM17,ROF92,CCZ13,CP16} for more detail).

{\it Bayesian inference}.
Estimating $\vect x$ (or $\vect a$) in problem \eqref{eqn:y} can be addressed in the Bayesian statistical framework \cite{R01}. 
Using Bayes' theorem we obtain the posterior distribution
\begin{equation}
p(\vect x | \vect y) = {p(\vect y | \vect x) p(\vect x)}/{\int_{\mathbb{R}^N}p(\vect y | \vect x) p(\vect x) {\rm d} \vect x},
\end{equation}
which models our knowledge about $\vect x$ after observing $\vect y$ given prior information, and
where $p(\vect y | \vect x)$ and $p(\vect x)$ respectively represent the likelihood function and prior distribution
(used to regularise the original problem, reduce uncertainty, and improve estimation results) of $\vect x$.
Using Bayes' theorem to model $\vect a$, $p(\vect a | \vect y)$ is given analogously. The general forms of 
the likelihood function and prior distribution considered are 
$p(\vect y|\vect x) \propto {\rm exp}(-g_{\vect y}(\vect x))$ and $p(\vect x) \propto {\rm exp}(-\mu f(\vect x))$,
respectively, where $\mu$ represents the so-called regularisation parameter which controls the strength of the prior information. 
The classical forms used are $g_{\vect y}(\vect x) = \|\vect y - \bm{\mathsf{\Phi}} \vect x \|_q^q /2\sigma^2$, 
$f(\vect x) = \|\bm{\mathsf{\Psi}}^\dagger \vect x\|_s$, where $q, s\ge 0$ (often $s =1$ is used), and $\sigma$ represents 
the standard deviation of the noise level.

{\it MAP estimation}.
Sampling the full posterior $p(\vect x | \vect y)$ or $p(\vect a | \vect y)$ by e.g. MCMC methods is difficult when 
high dimensionality is involved \cite{CPM17,CPM17b,DMP16,M15,PSCPTHM16}. Instead, Bayesian 
estimators that summarise $p(\vect x | \vect y)$ or $p(\vect a | \vect y)$ are often computed. 
A particularly popular choice is the MAP estimator, given by
\begin{equation}\label{eqn:un-af}
{\vect x}^*_{\mu}  = \argmin_{\vect x \in \mathbb{R}^{N}} \big\{ \mu f(\vect x) +  g_{\vect y}(\vect x) \big\}.
\end{equation}
In the rest of this article we assume $f(\vect x)$ and $g_{\vect y}(\vect x)$ are closed convex functions which are not necessarily
differentiable, and the objective functional \eqref{eqn:un-af} is computationally tractable with respect to $\vect x$ given the value of $\mu$. 
For further details about MAP estimation see, e.g., \cite{Pereyra:2016b}.

A main computational advantage of the MAP estimator \eqref{eqn:un-af}
is that it can be computed very efficiently, even in high dimensions, by using convex optimisation algorithms (e.g. \cite{CP10,CP16,Green2015}). 
However, since MAP estimation results in a single point estimator, we lose uncertainty information that sampling approaches like MCMC methods can provide \cite{CPM17}. 
But as shown in \cite{M16}, it is possible to approximately quantify the uncertainties associated with $\vect x | \vect y$ by leveraging recent results in the theory of probability concentration.

{\it Optimisation algorithms}.
There are many convex minimisation methods that can be used to
solve the MAP estimation problems with form \eqref{eqn:un-af} efficiently, such
as forward-backward splitting, Douglas-Rachford splitting, primal-dual, or alternating
direction method of multipliers (see \cite{CP10,CFNSS15,BPCPE10} and references therein for more detail).
Furthermore, algorithmic structures that allow computations
to be highly distributed and parallelised have also been developed, see e.g. \cite{car14,OCRMTPW16,PJM18,BPCPE10}.
Note that to solve problem \eqref{eqn:un-af}, the regularisation parameter $\mu$ should either be given
beforehand or be estimated from $\vect y$.

Selecting the value of the regularisation parameter $\mu$ in \eqref{eqn:un-af} correctly is crucial, as this value impacts strongly the estimation results \cite{MBF15}. Setting appropriate values for $\mu$ by hand is generally difficult, and fortunately there are now several different kinds of Bayesian and non-Bayesian methods to select $\mu$ automatically \cite{MBF15,OBF09,PBC09, DVPF14}. 
In the following, we briefly recall the Bayesian method \cite{MBF15} (coming from a hierarchical Bayesian model and
using joint MAP estimation), which we adopt for our 
UQ strategies. The resulting iteration formula is given by
\begin{align}\label{eqn:para-sel}
\begin{split}
{\vect x}^{(i)} &= \argmin_{\vect x \in \mathbb{R}^{N}} \big\{ \mu^{(i-1)} f(\vect x) +  g_{\vect y}(\vect x) \big\}, \\
\mu^{(i)} &= (N/k + \gamma -1)/(f({\vect x}^{(i)}) + \beta),
\end{split}
\end{align}
where $\gamma, \beta$ are fixed parameters (default values are 1), and $k$ is related to 
the definition of $f$ and is fixed as well (e.g. $k = 1$ when $f$ is $\ell_1$ norm);
refer to \cite{MBF15} for more detail.

\section{Proposed Uncertainty Quantification Methods}\label{sec:method}
As discussed previously, MCMC methods sampling the full posterior $p(\vect x | \vect y)$ [or $p(\vect a | \vect y)$] 
to quantify uncertainties is computationally expensive when high dimensionality is involved. 
Instead, Bayesian credible regions can be estimated approximately based on the MAP estimator that 
summarises $p(\vect x | \vect y)$ [or $p(\vect a | \vect y)$], which does scale efficiently to high-dimensional
settings \cite{M16, CPM17b}. 

\begin{figure}[!htb] 
\begin{center}
    \begin{tabular}{c}

\begin{tikzpicture}[>=stealth,every node/.style={shape=rectangle,draw,rounded corners},]
    \node (c0){};
    \node (c1) [fill=red!30]{Observations: $\vect y$};
    \node (c00) [fill=blue!20, below = 0.35cm of c1, node distance=2cm, text width=6.3cm,align=center]{Inverse problem: form objective functional };
    \node (c01) [fill=blue!20, below = 0.35cm of c00, node distance=2cm, text width=7.5cm,align=center]{Regularisation parameter selection (automatically) };
    \node (c2) [fill=blue!20, below = 0.35cm of c01, node distance=2cm, text width=5.5cm,align=center]{MAP image/signal estimation: ${\vect x}^*_{\mu}$ };
    \node (c5) [fill=green!10, below = 0.4cm of c2,text width=2.0cm,align=center]{Approximate local credible intervals: $({\vect \xi}_-, {\vect \xi}_+)$ };
    \node (c3) [fill=green!10, left of = c5, node distance=2.8cm, text width=2.0cm,align=center]{Approximate HPD credible regions: $C^{\prime}_{\alpha}$  };

    \draw[->] (c1) -- (c00);
    \draw[->] (c00) -- (c01);
    \draw[->] (c01) -- (c2);
    \draw[->] (c2) -- (c3);
    \draw[->] (c3) -- (c5);
    \draw[->] (c2) -- (c5);
    \draw[->] (c1) --++  (-4.1,0)  |-  (c3.west);
\end{tikzpicture}
    \end{tabular}
\end{center} 
\caption{UQ methodology based on MAP estimation for image/signal processing. 
The light green areas at the bottom show the types of UQ developed. 
}\label{Fig-flow_chart}
\end{figure}
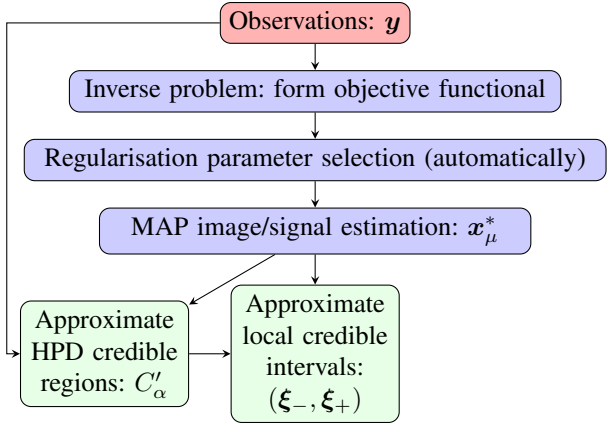

The diagram in Fig. \ref{Fig-flow_chart} shows the main procedures of our proposed UQ methodology based on MAP estimation.
Firstly, the objective functional related to a given inverse problem is formed, followed by 
estimation of $\mu$ {(e.g. using the method in \cite{MBF15}, shown in \eqref{eqn:para-sel})} and MAP estimation of $\vect x$ 
using convex optimisation techniques (which can scale to high-dimensional problems
where MCMC methods struggle). Then, various forms of UQ, e.g. those described in \cite{CPM17b}, are performed.
Here we restrict our attention on local credible interval at pixel level ({\it cf.} error bars).

The first step in our UQ pipeline is to compute the HPD region $C_{\alpha}$ with level $(1-\alpha)$, defined by 
$C_{\alpha} := \{\vect x: \mu f(\vect x) + g_{\vect y}(\vect x) \le \gamma_{\alpha}\}$ \cite{R01}, with $\gamma_{\alpha}$ 
such that  
\begin{equation}
p (\vect x \in C_{\alpha} | \vect y) = \int_{\vect x \in \mathbb{R}^N} p(\vect x | \vect y) \mathbb{1}_{C_{\alpha}} {\rm d} \vect x = 1-\alpha\,.
\end{equation}
Computing $C_{\alpha}$ exactly is often not possible when $N = \textrm{dim}(\vect x)$ is large. However, 
from \cite{M16}, we can have an approximation of $C_{\alpha}$, say $C^{\prime}_{\alpha}$, obtained by approximating $\gamma_{\alpha}$ by $\gamma^{\prime}_{\alpha}$ given by
\begin{equation}\label{eqn:gamma}
{\gamma}^{\prime}_{\alpha} = \mu f({\vect x}^*_{\mu}) +  g_{\vect y}({\vect x}^*_{\mu}) + \sqrt{16\log(3/\alpha)}\sqrt{N} + N.
\end{equation}
{This approximation was motivated from recent results in information theory in terms of a probability
concentration inequality, and is generally accurate for large $N$ (refer to \cite{M16,CPM17b} for more details).}

The second step in the pipeline is to use $C^{\prime}_{\alpha}$ to compute local credible interval for pixels and superpixels. This is a form 
of Bayesian UQ that we use as a means for visualising uncertainty spatially at different scales \cite{CPM17b}.
Let $\Omega = \cup_{i}\Omega_i $ be a partition of the image/signal domain $\Omega$ into subsets or \emph{superpixels} $\Omega_i$ 
satisfying $\Omega_i\cap\Omega_j = \emptyset, i\neq j$. Then a local credible interval $({\xi}_{-, \Omega_i}, {\xi}_{+,  \Omega_i})$
for region $\Omega_i$ is defined by 
{\small
\begin{align}
{\xi}_{-,  \Omega_i} & \!=\! \min_{\xi}\left \{ \xi | \mu f({\vect x}_{i,\xi}) \!+\! g_{\vect y}({\vect x}_{i,\xi}) \le {\gamma}^\prime_{\alpha}, \forall \xi \in [0, +\infty )  \right \},  
\label{eqn:cr-local-sp-i-l}\\
{\xi}_{+,  \Omega_i} & \!=\! \max_{\xi}\left \{ \xi | \mu f({\vect x}_{i,\xi}) \!+\! g_{\vect y}({\vect x}_{i,\xi}) \le {\gamma}^\prime_{\alpha}, \forall \xi \in [0, +\infty )  \right \},  
\label{eqn:cr-local-sp-i-h}
\end{align} }
\vspace{-0.15in}

\noindent {where ${\vect x}_{i,\xi} = {\vect x}_{\mu}^* \vect \zeta_{\Omega\setminus\Omega_i} + \xi \vect \zeta_{\Omega_i}$, 
$\vect \zeta_{\Omega_i} \in \mathbb{R}^N$ is the index operator on $\Omega_i$ with value 1 for pixels in $\Omega_i$ otherwise 0.}
{Note that ${\xi}_{-,  \Omega_i}$ and ${\xi}_{+,  \Omega_i}$ are actually the values that saturate the HPD credible
region $C^{\prime}_{\alpha}$ from above and from below at $ \Omega_i$. }
Then the local credible interval $({\vect \xi}_-, {\vect \xi}_+)$ for the whole image/signal is obtained by gathering all the 
$({\xi}_{-, \Omega_i}, {\xi}_{+,  \Omega_i}), \forall i$, i.e.,
\begin{equation}
{\vect \xi}_- = \sum_i {\xi}_{-, \Omega_i} \vect \zeta_{\Omega_i}, \quad {\vect \xi}_+ = \sum_i {\xi}_{+, \Omega_i} \vect \zeta_{\Omega_i}.
\end{equation}

We hereby briefly clarify the distinctions of this work from \cite{CPM17b}. 
Firstly, we now concern the UQ strategies in general image/signal processing problems instead of just a 
special application in RI imaging in \cite{CPM17b}. Secondly, 
here we adjust $\mu$ automatically, but \cite{CPM17b} assumes $\mu$ is known beforehand.
Finally, we consider the over-complete bases $\bm{\mathsf{\bm{\mathsf{\Psi}}}}$ (such as SARA \cite{CMW12,CMVTW13}) and explore their influence in 
UQ with synthesis and analysis priors, which is not considered in \cite{CPM17b}.

\section{Experiments}\label{sec:experiment}
\addtolength{\tabcolsep}{-\tabL}
\begin{figure}[!htb]
\begin{center}
\begin{tabular}{cc}
\includegraphics[trim={{.15\linewidth} {.07\linewidth} {.065\linewidth} {.16\linewidth}}, clip, width=0.39\linewidth, height = 0.35\linewidth]{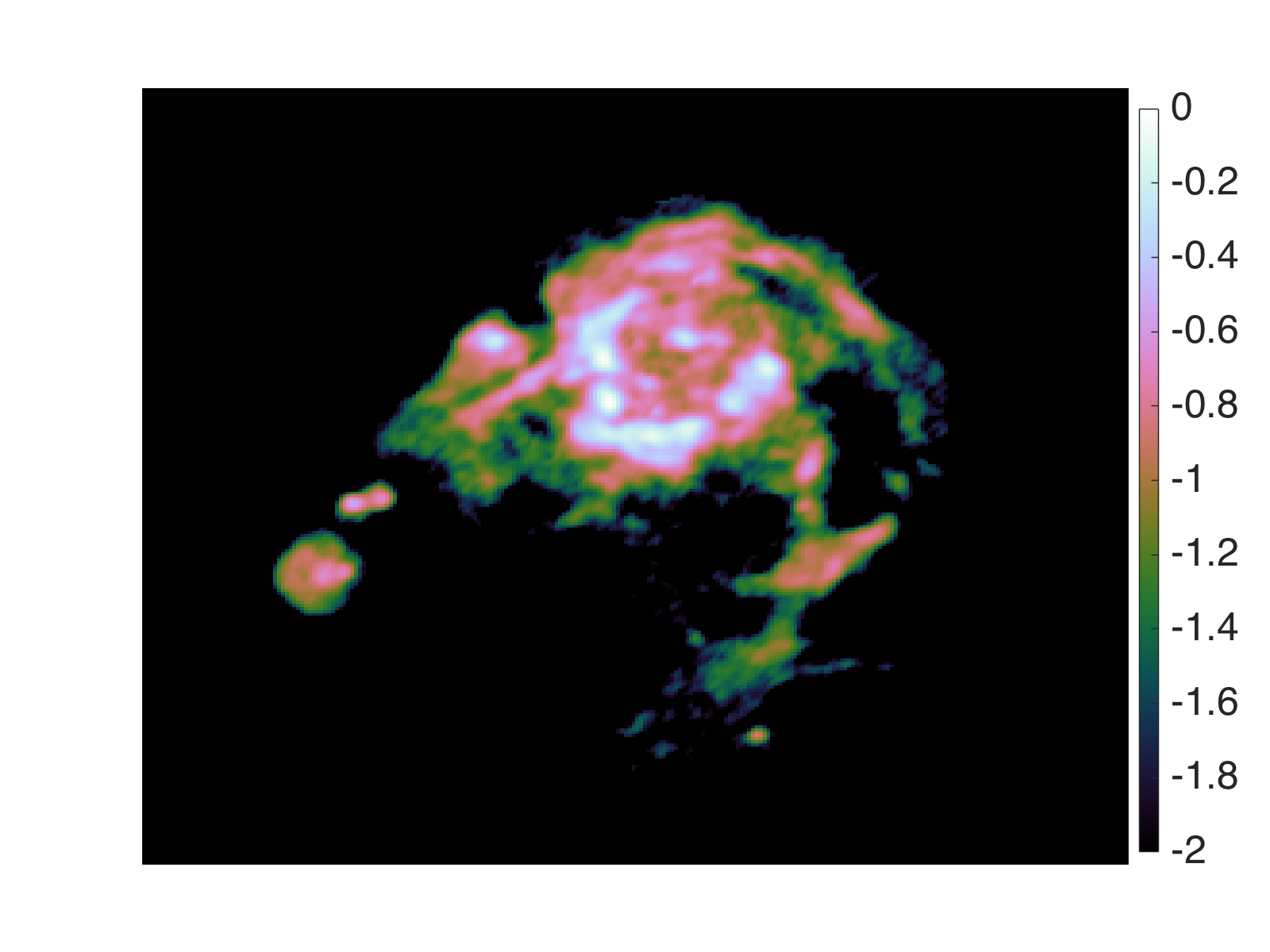} &
\includegraphics[trim={{.15\linewidth} {.07\linewidth} {.065\linewidth} {.16\linewidth}}, clip, width=0.39\linewidth, height = 0.35\linewidth]{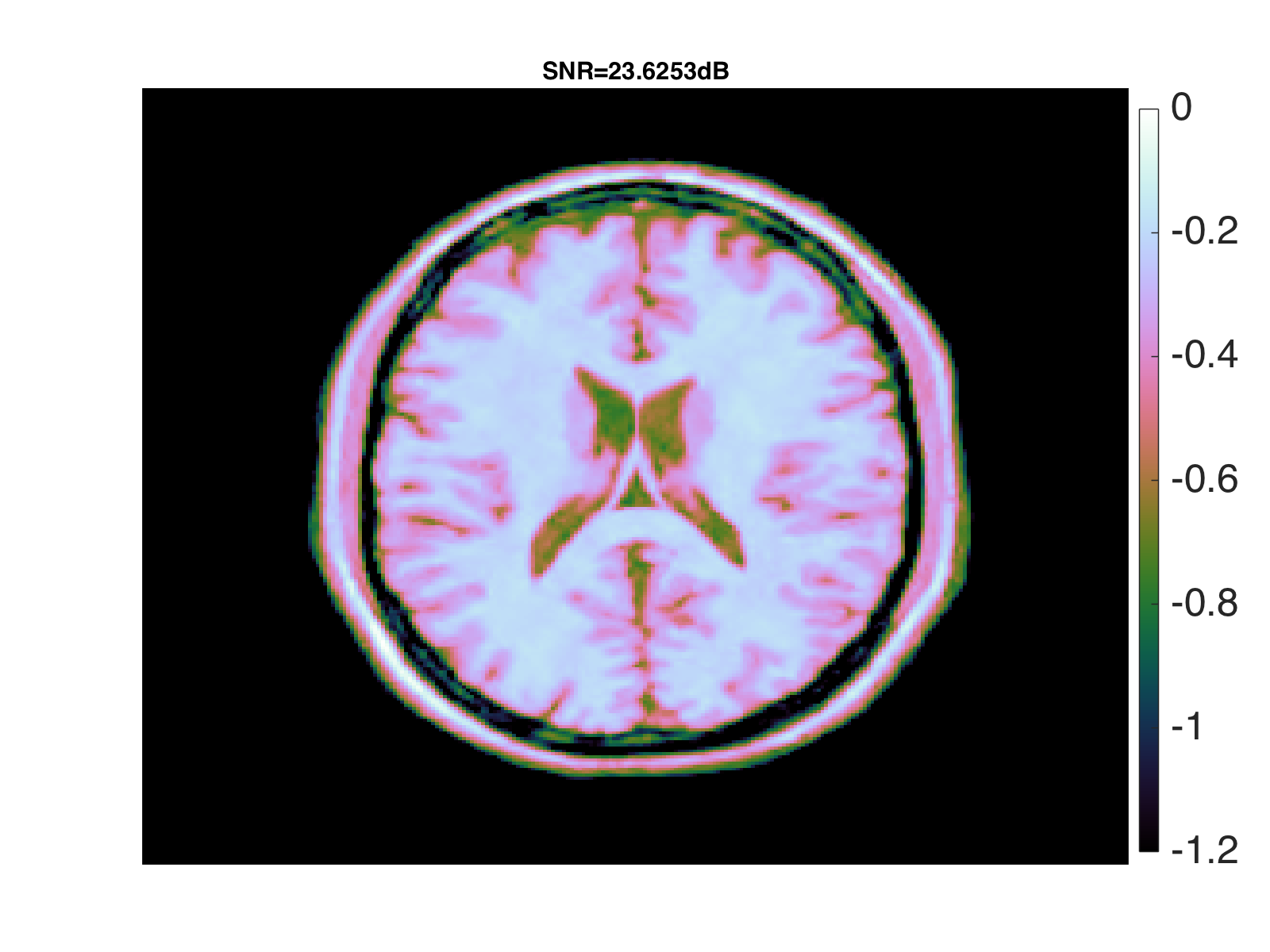} 
\end{tabular}
\end{center}
\caption{Test images in applications of RI imaging and medical imaging; shown in ${\tt log}_{10}$ scale. Left: RI image M31; right: MRI brain image.
}\label{Fig-test-image}
\end{figure}
\addtolength{\tabcolsep}{\tabL}

\begin{table}
\begin{center}
  \caption{Reconstruction quality in SNR and the automatically computed parameter $\mu$ with
  orthonormal basis and SARA library, corresponding 
  to synthesis and analysis priors. 	
  }      
 \label{tab:snr-u}
\begin{tabular}{ccccc}
\toprule  
 \multirow{1}{*}{Image}    & \multirow{1}{*}{Library/basis} &  Synthesis  &  Analysis  &   \multirow{1}{*}{$\mu$}  
\\ \toprule 
\multirow{2}{*}{M31 }  &  Orthonormal  & $25.04$ & $25.04$  & 196 \\ 
&  SARA  & $23.66$ & $31.09$  & 65 \\ \hline
\multirow{2}{*}{Brain }   & Orthonormal  & $19.06$ & $19.06$ & 33 \\ 
& SARA & $19.89$ & $23.63$  & 11 \\ \bottomrule
\end{tabular}
\end{center}
\end{table}

\addtolength{\tabcolsep}{-\tabL}
\begin{figure}[!htb]
\begin{center}
\begin{tabular}{cc}
\includegraphics[trim={{.09\linewidth} {.07\linewidth} {.025\linewidth} {.03\linewidth}}, clip, width=0.51\linewidth, height = 0.43\linewidth]{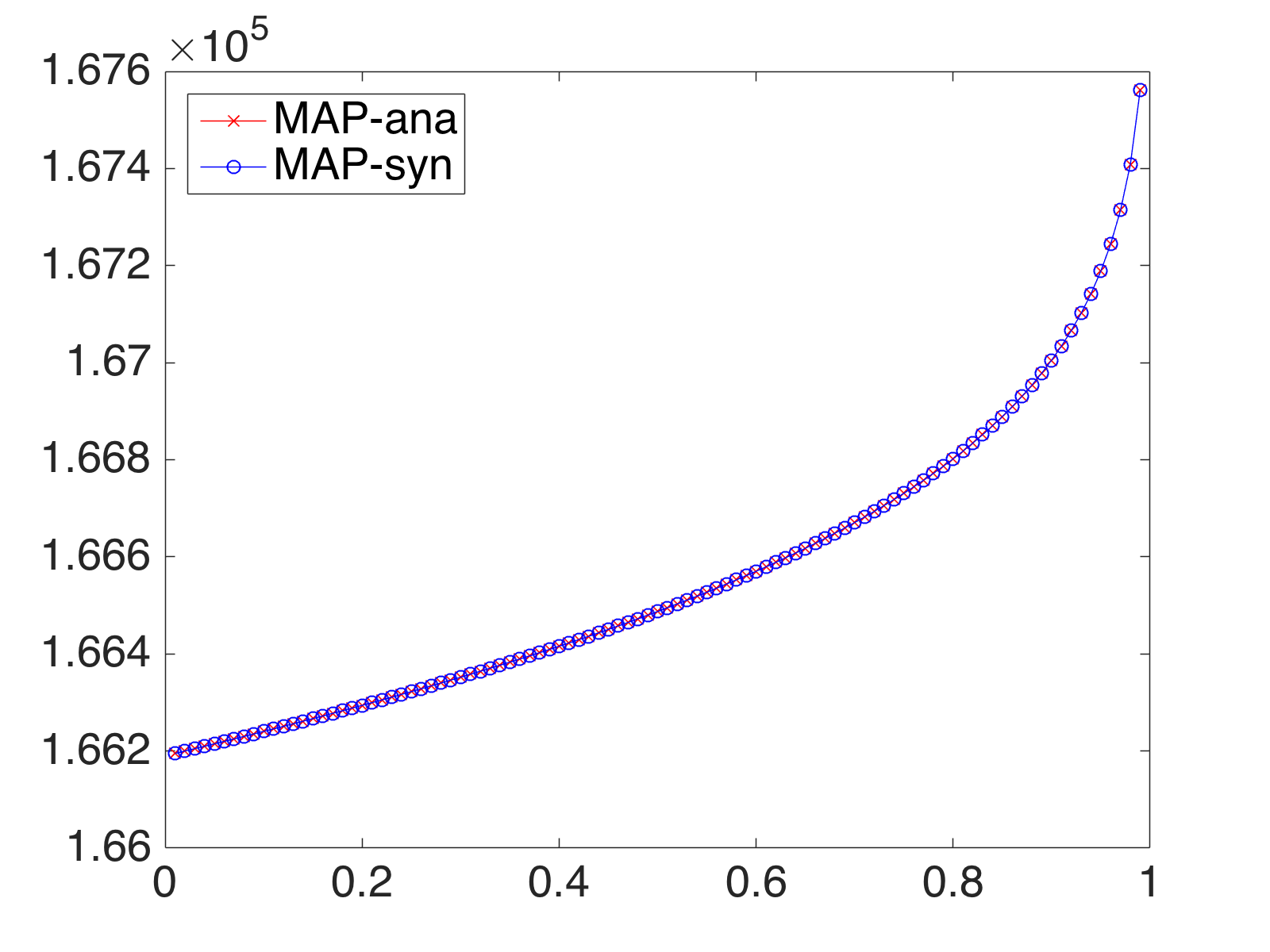} 
\put(-76,-6){\small $1-\alpha$} & 
\includegraphics[trim={{.12\linewidth} {.07\linewidth} {.025\linewidth} {.03\linewidth}}, clip, width=0.51\linewidth, height = 0.43\linewidth]{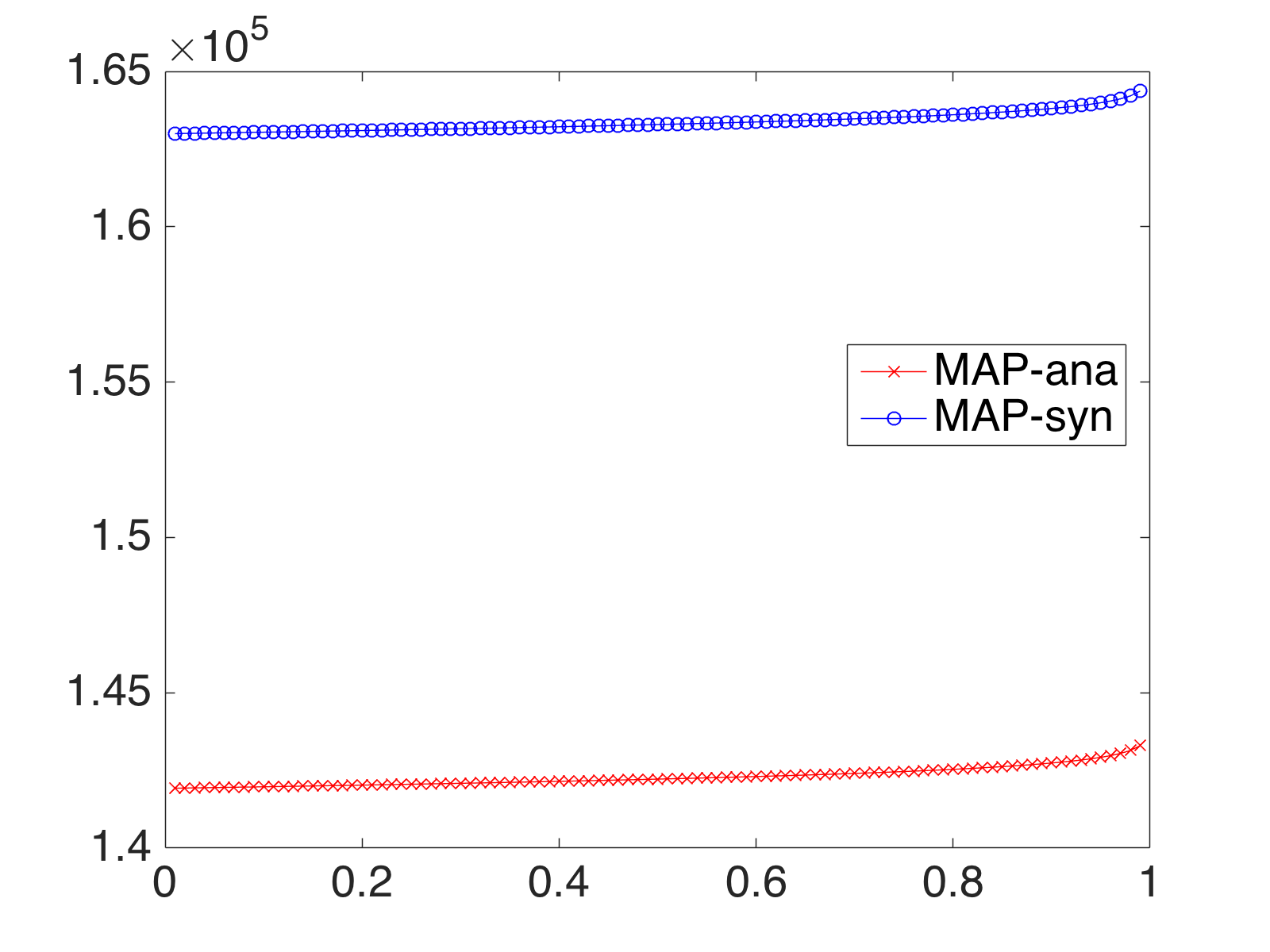} 
\put(-76,-6){\small $1-\alpha$}
\end{tabular}
\end{center}
\caption{HPD credible region. Plots on the left and right are the results using orthonormal basis and SARA dictionary, respectively. MRI brain image is used as an example here
(results for RI image M31 are similar). 
}\label{Fig-HPD}
\end{figure}
\addtolength{\tabcolsep}{\tabL}

\addtolength{\tabcolsep}{-\tabL}
\begin{figure*}[!htb]
\begin{center}
\begin{tabular}{ccccc}
\includegraphics[trim={{.15\linewidth} {.07\linewidth} {.022\linewidth} {.074\linewidth}}, clip, width=0.19\linewidth, height = 0.17\linewidth]{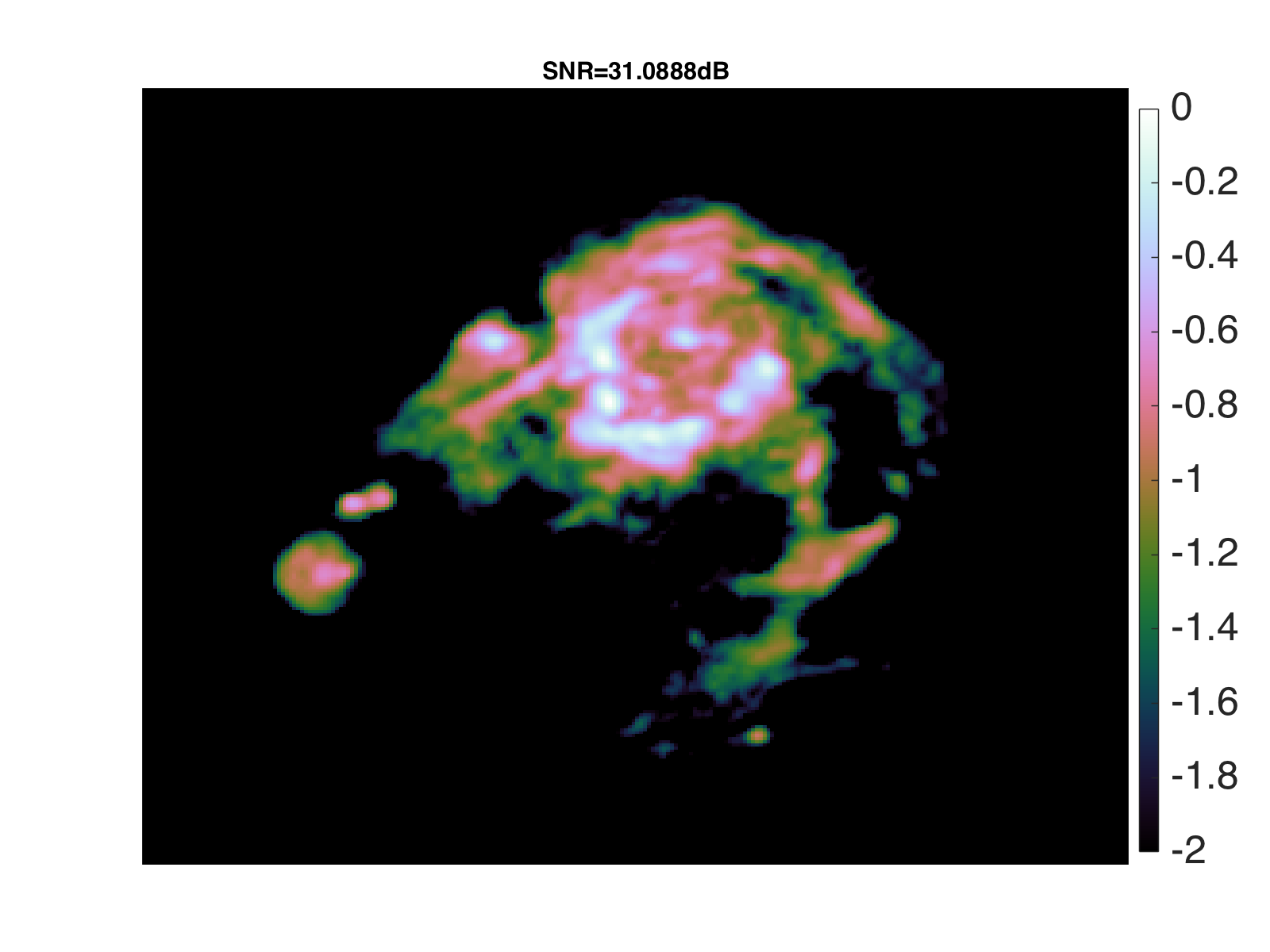}  &
\includegraphics[trim={{.15\linewidth} {.07\linewidth} {.022\linewidth} {.074\linewidth}}, clip, width=0.19\linewidth, height = 0.17\linewidth]{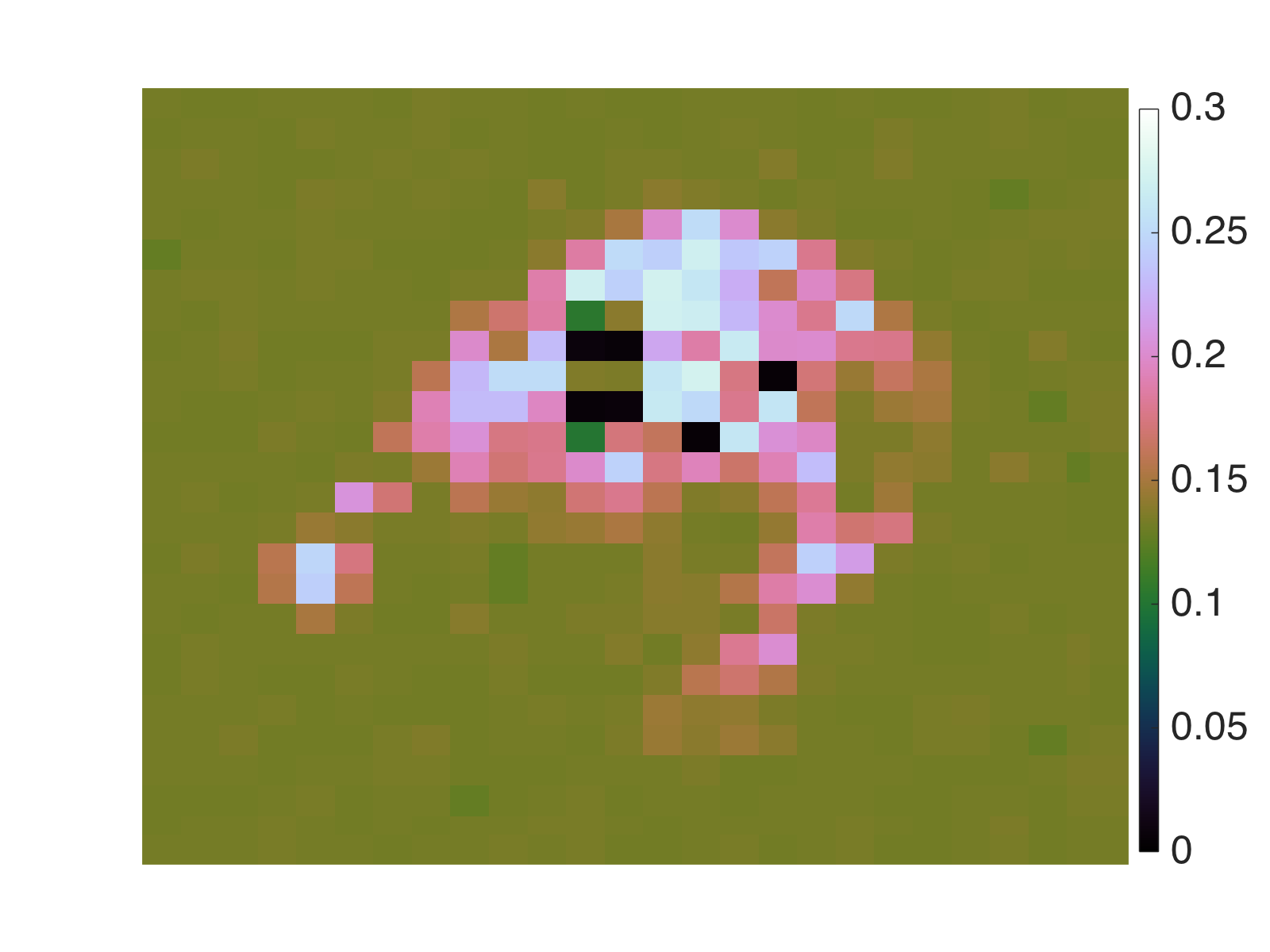}  &
\includegraphics[trim={{.15\linewidth} {.07\linewidth} {.022\linewidth} {.074\linewidth}}, clip, width=0.19\linewidth, height = 0.17\linewidth]{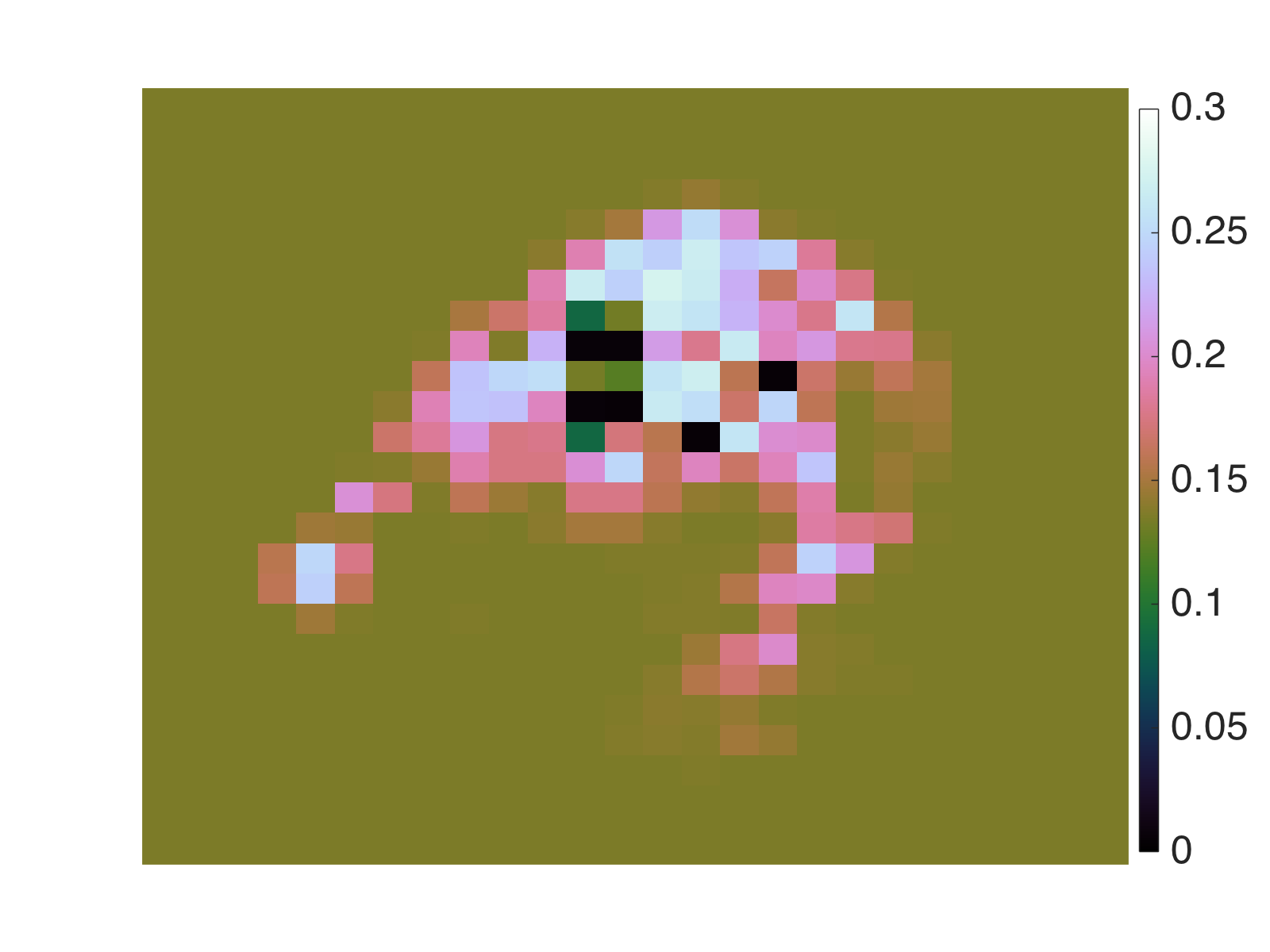}  &
\includegraphics[trim={{.15\linewidth} {.07\linewidth} {.022\linewidth} {.074\linewidth}}, clip, width=0.19\linewidth, height = 0.17\linewidth]{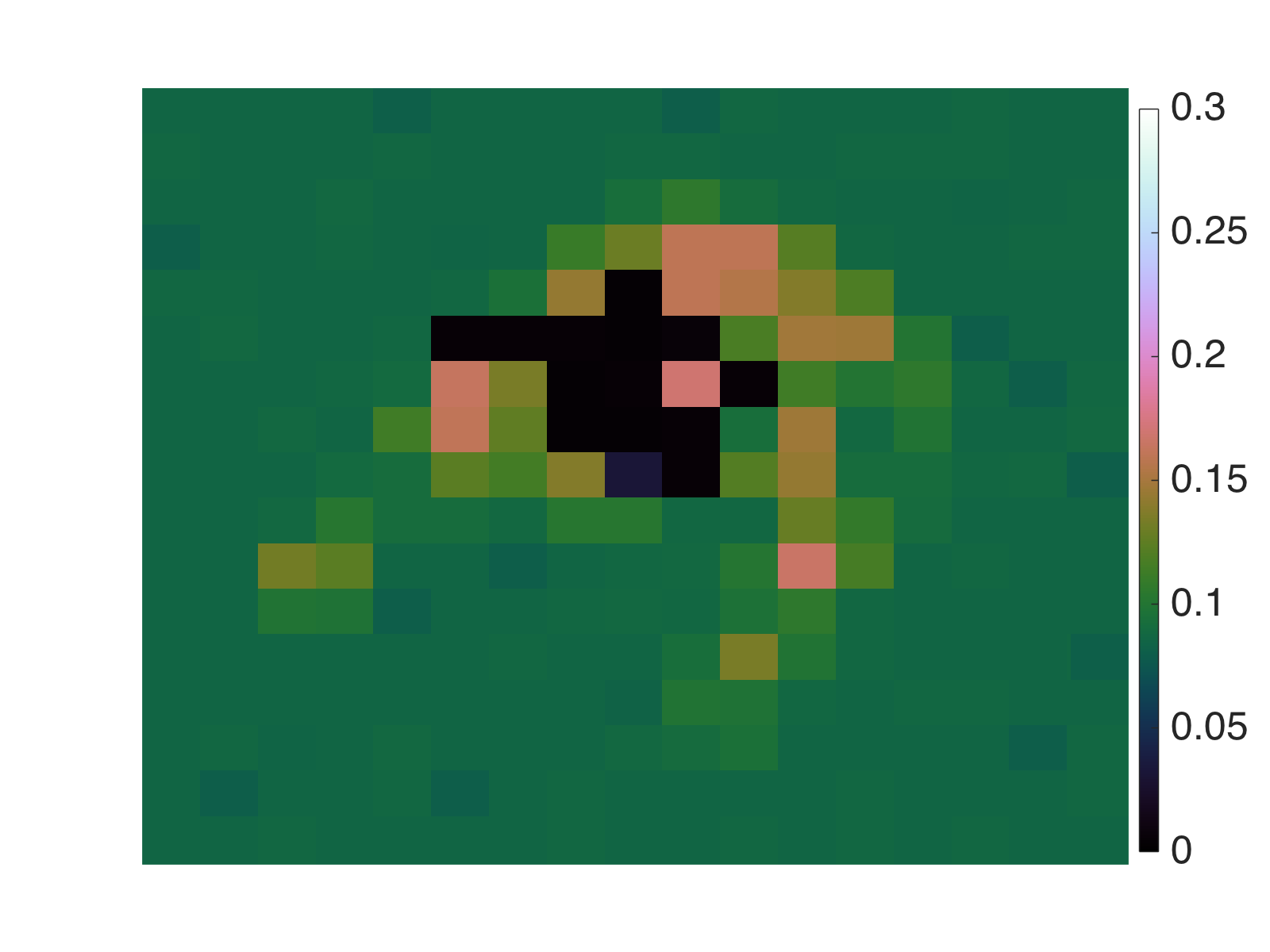}  &
\includegraphics[trim={{.15\linewidth} {.07\linewidth} {.022\linewidth} {.074\linewidth}}, clip, width=0.19\linewidth, height = 0.17\linewidth]{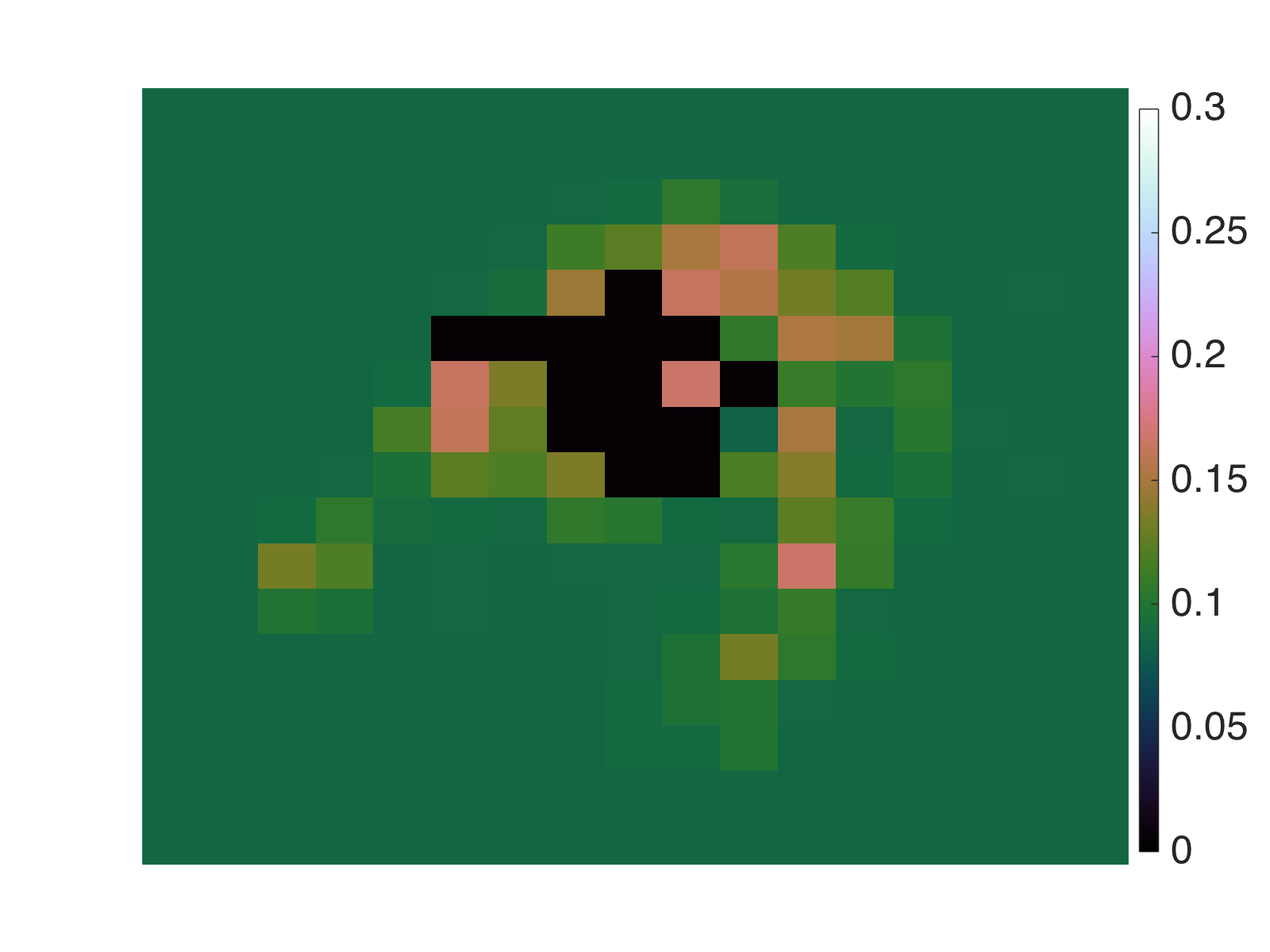}  \\
\includegraphics[trim={{.15\linewidth} {.07\linewidth} {.022\linewidth} {.074\linewidth}}, clip, width=0.19\linewidth, height = 0.17\linewidth]{figs/MRI_result_ana_sara_1.png}  &
\includegraphics[trim={{.15\linewidth} {.07\linewidth} {.022\linewidth} {.074\linewidth}}, clip, width=0.19\linewidth, height = 0.17\linewidth]{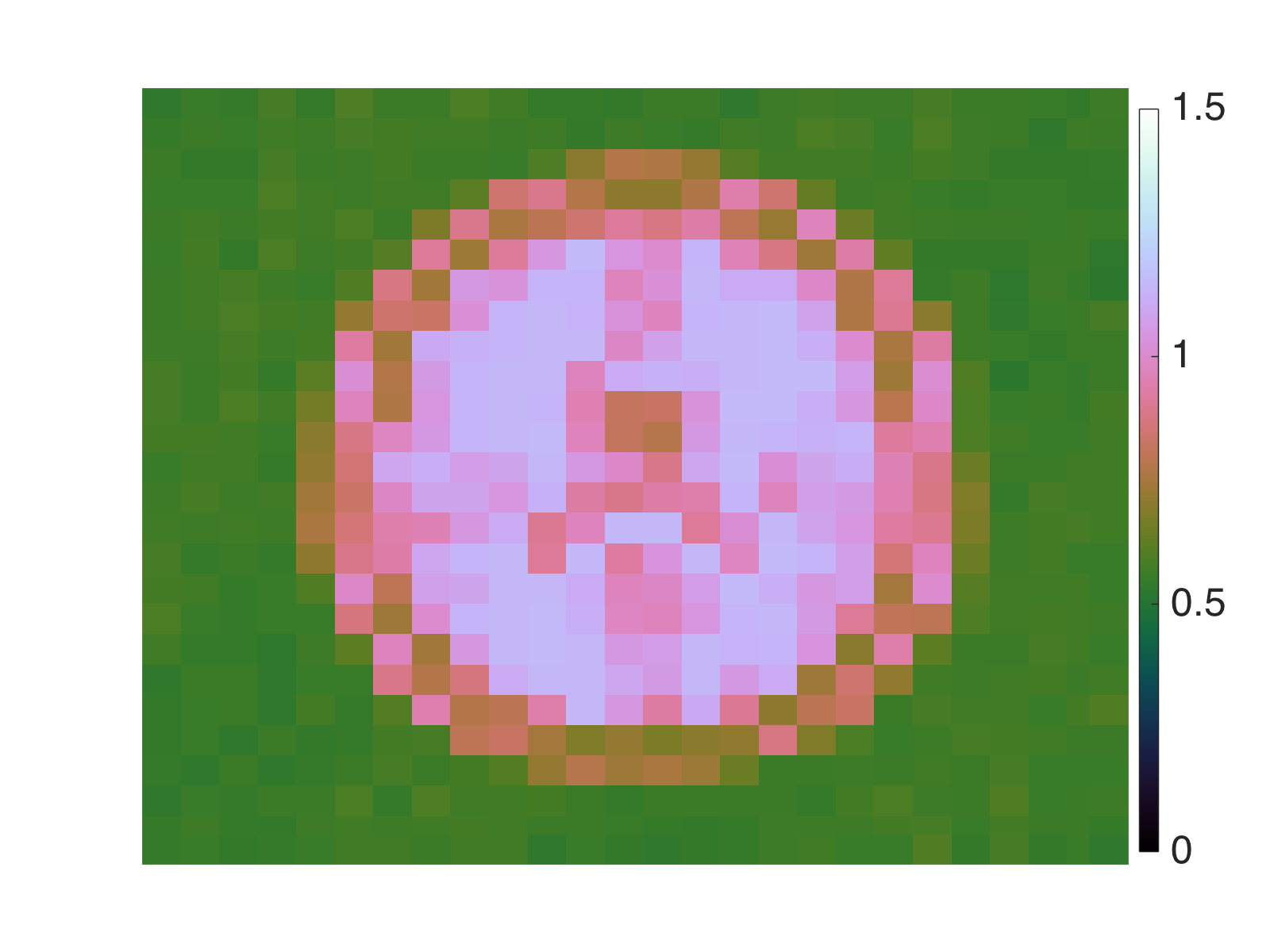}  &
\includegraphics[trim={{.15\linewidth} {.07\linewidth} {.022\linewidth} {.074\linewidth}}, clip, width=0.19\linewidth, height = 0.17\linewidth]{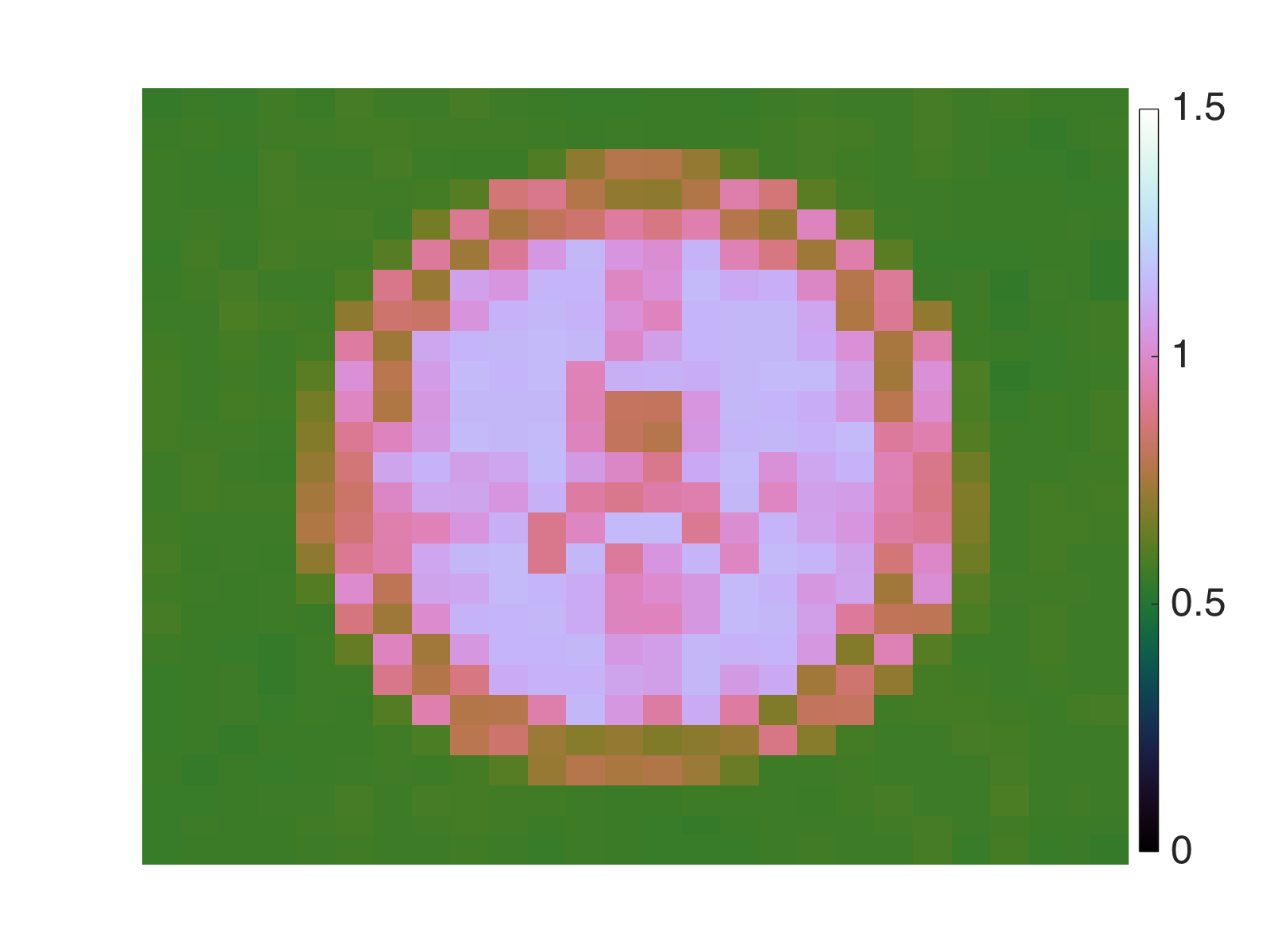}  &
\includegraphics[trim={{.15\linewidth} {.07\linewidth} {.022\linewidth} {.074\linewidth}}, clip, width=0.19\linewidth, height = 0.17\linewidth]{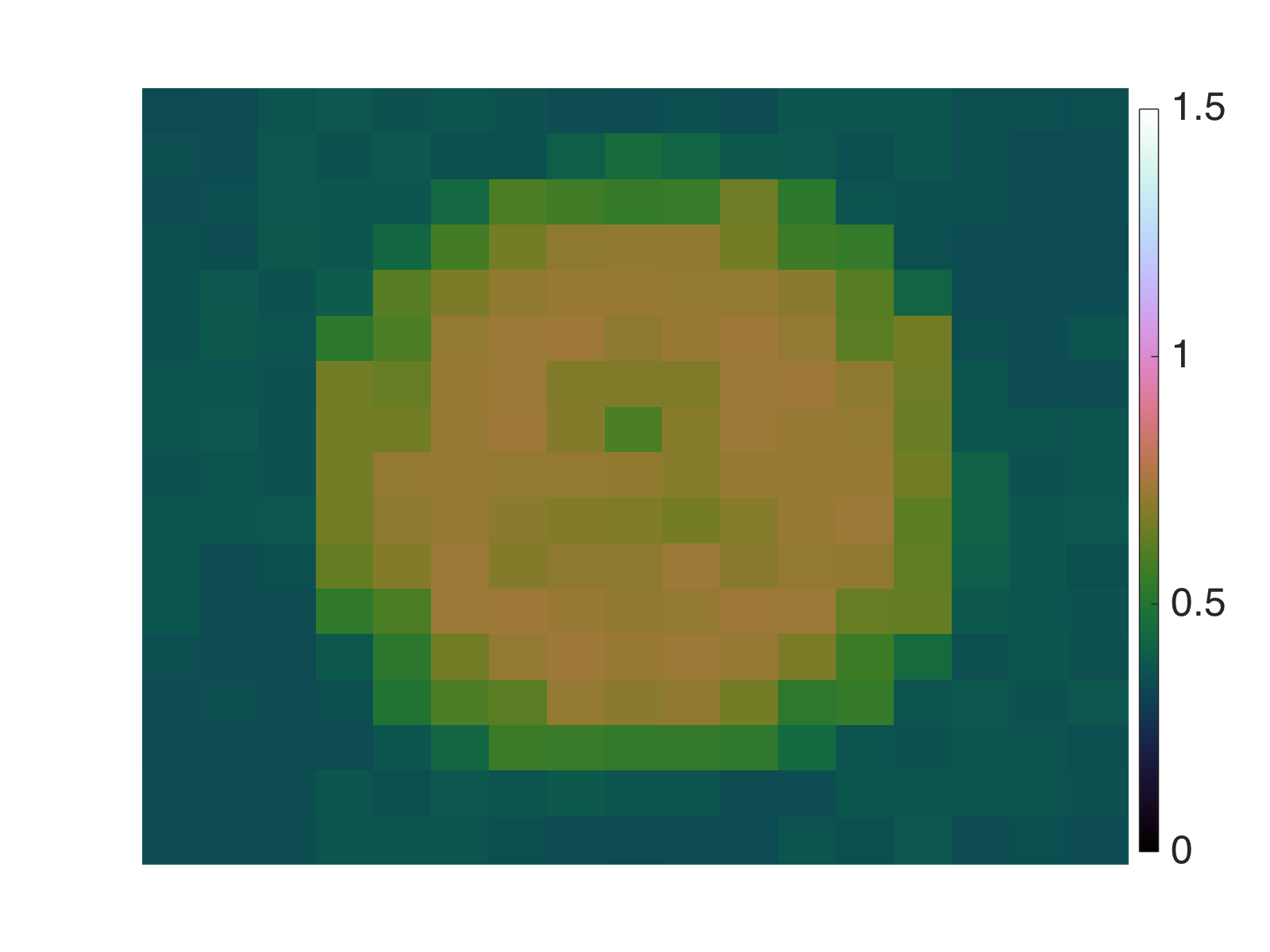}  &
\includegraphics[trim={{.15\linewidth} {.07\linewidth} {.022\linewidth} {.074\linewidth}}, clip, width=0.19\linewidth, height = 0.17\linewidth]{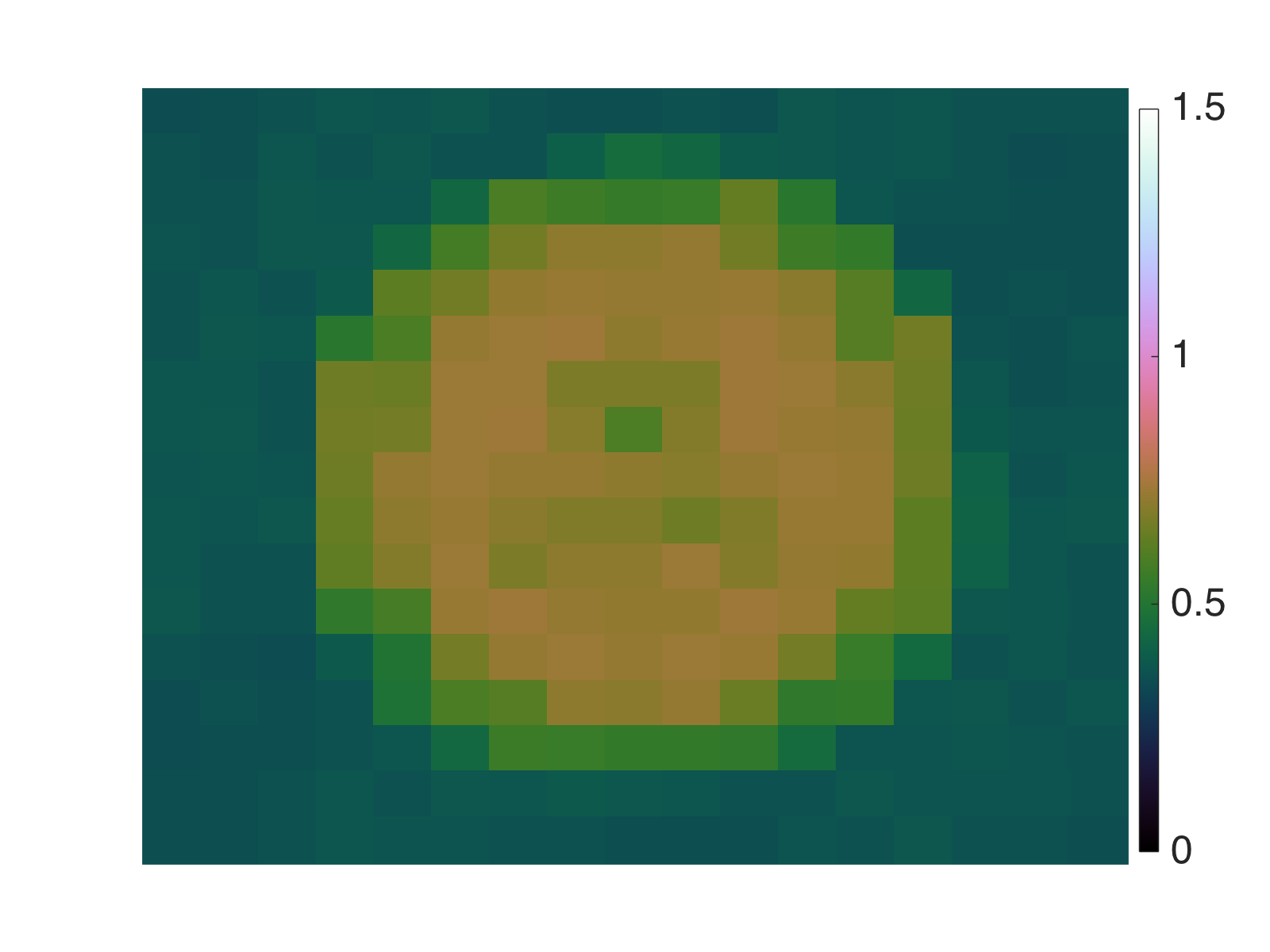}  \\
{\small  (a) ${\vect x}^*_{\mu}$ } & {\small (b) synthesis ($10\times 10$) } &{\small (c) analysis ($10\times 10$) } & 
{\small (d) synthesis ($15\times 15$)} &{\small (e) analysis ($15\times 15$)} 
\end{tabular}
\end{center}
\caption{Length of the local credible interval (99\% credible level) for RI image M31 (row one) and MRI brain image (row two) (size $256\times 256$). 
(a) is the computed point estimator for the objective functional equipped with analysis prior and SARA dictionary and is
 shown in ${\tt log}_{10}$ scale;  (b)--(c) and (d)--(e) are the length of the computed local credible interval obtained by equipping
SARA dictionary at grid scales of $10\times 10$ and $15\times 15$, respectively. 
}\label{Fig-lc-m31}
\end{figure*}
\addtolength{\tabcolsep}{\tabL}

\addtolength{\tabcolsep}{-\tabL}
\begin{figure}[!htb]
\begin{center}
\begin{tabular}{cc}
\includegraphics[trim={{.15\linewidth} {.09\linewidth} {.08\linewidth} {.1\linewidth}}, clip, width=0.48\linewidth, height = 0.35\linewidth]{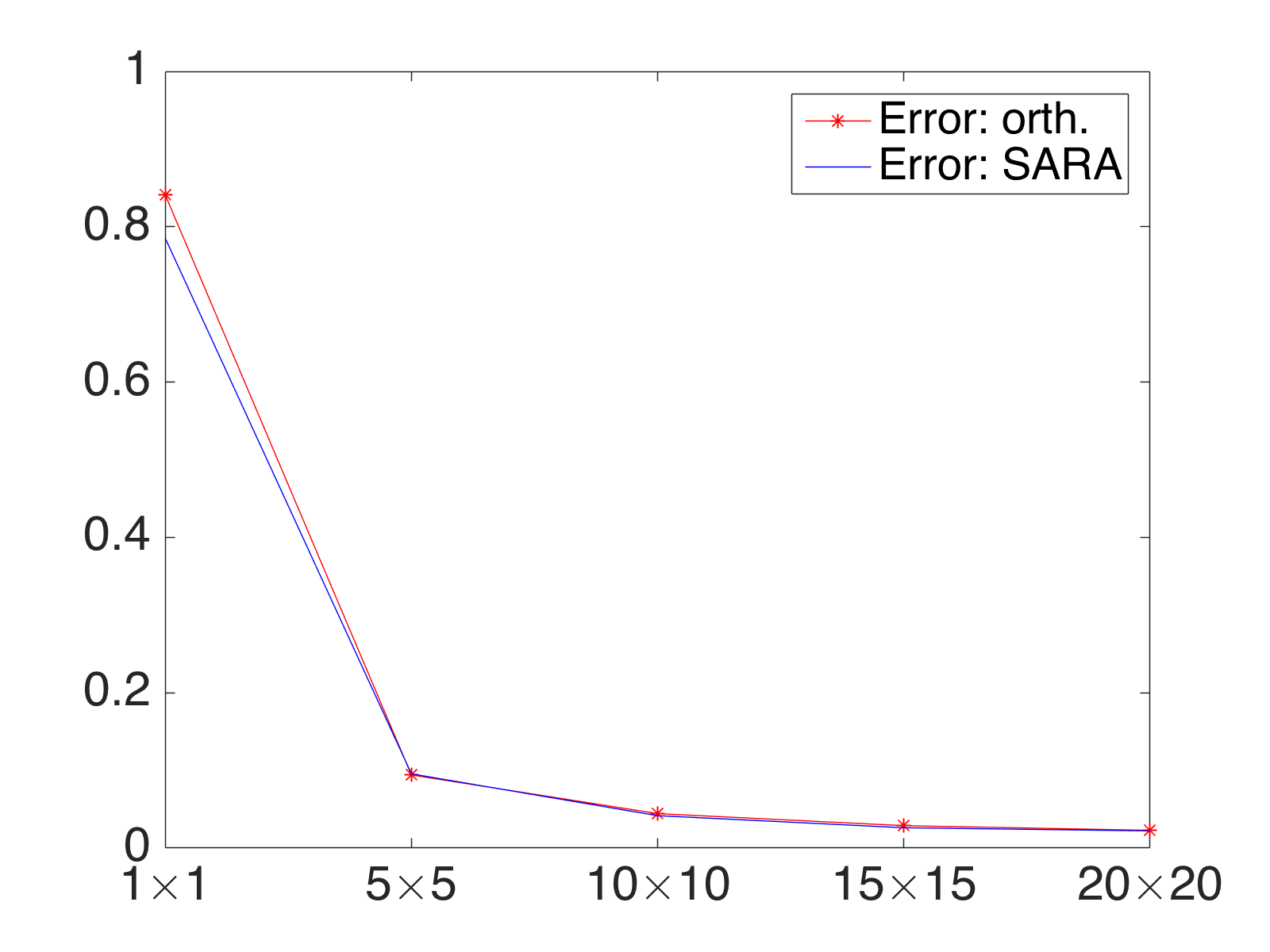} 
 \put(-128,18){\rotatebox{90}{\small relative error} } \put(-112,75){ {\footnotesize Test image: M31} } &
\includegraphics[trim={{.15\linewidth} {.09\linewidth} {.08\linewidth} {.1\linewidth}}, clip, width=0.48\linewidth, height = 0.35\linewidth]{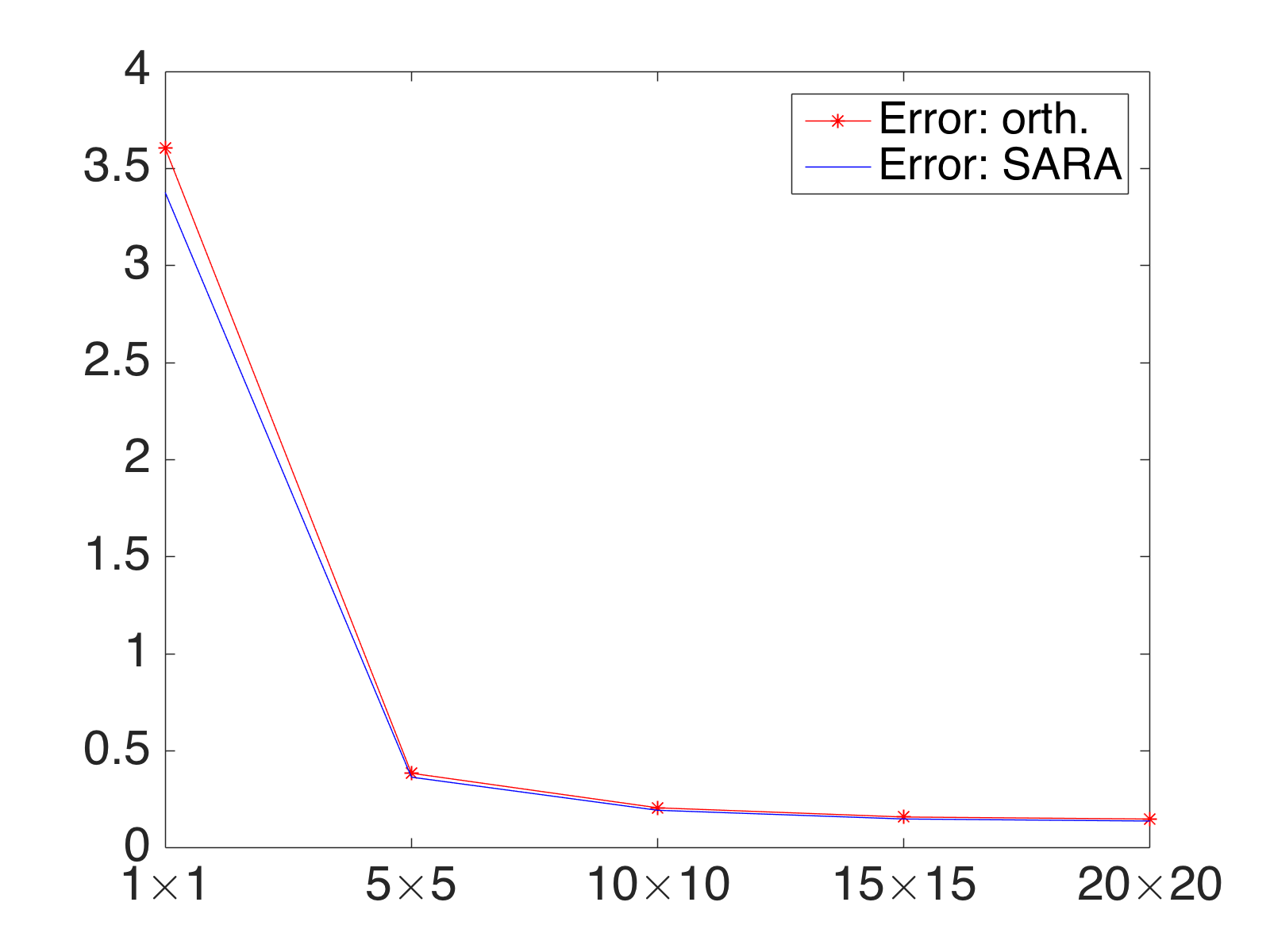} 
\put(-142,-10){\small grid scale}   \put(-112,75){ {\footnotesize Test image: Brain} }
\end{tabular}
\end{center}
\caption{Average relative error (over all pixels) of the length of the local credible interval (99\% credible level) computed by MAP estimation,
where the dividend of the error is the absolute difference of the length of the local credible interval computed by MAP estimation and Px-MALA, and the divisor is the difference 
of the maximum and minimum values of the clean image. 
The analysis prior is used here and the test images are the RI image M31 (left plot) and MRI brain image (right plot). The red line with asterisk and the blue line are the errors regarding 
the orthonormal basis and SARA dictionary, respectively. 
}\label{Fig-error}
\end{figure}
\addtolength{\tabcolsep}{\tabL}

We now illustrate the proposed UQ methodology with a canonical image processing problem --
image reconstruction with $\ell_1$ prior. 
The observations (noisy measurement) ${\vect y} = \bm{\mathsf{\Phi}} {\vect x} + \vect n \in \mathbb{C}^{M}$
where $M = N/10$ is used, and $\bm{\mathsf{\Phi}}$ is constructed using Fourier transform followed by a downsampling mask. 
We consider both the analysis and synthesis priors, i.e., $f(\vect x) = \|\bm{\mathsf{\Psi}}^\dagger {\vect x}\|_1$
and $f(\vect a) = \|{\vect a}\|_1 $, respectively. Accordingly, the likelihood functions are 
set to $g_{\vect y} (\vect x) = \|{\vect y}-\bm{\mathsf{\Phi}} {\vect x}\|_2^2/2\sigma^2$ and 
$g_{\vect y} (\vect a) = \|{\vect y}-\bm{\mathsf{\Phi}}\bm{\mathsf{\Psi}} {\vect a}\|_2^2/2\sigma^2$,
where $\sigma = \|\vect x^*\|_{\infty} 10^{{\rm SNR}/20}$ with SNR (signal to noise ratio) set to 30.
In contrast to the work in \cite{CPM17,CPM17b}, which only used an orthonormal basis for $\bm{\mathsf{\Psi}}$ (Daubechies 8 [DB8]), here
we also consider an over-complete basis.  
Note that an over-complete basis can lead to difference between the analysis and synthesis priors,
and therefore is worth investigating. For comparison, we use DB8 wavelets for
the orthonormal basis and the over-complete SARA dictionary, consisting of a concatenation of nine bases (DB1--DB8 plus Dirac basis) \cite{CMW12}.

The numerical experiments performed in this article were run on a Macbook laptop with an i7 Intel CPU and
memory of 16 GB, running MATLAB R2015b.
We report experiments with two widely used test data sets -- one is M31 (Fig. \ref{Fig-test-image} left) in RI imaging \cite{CPM17,CPM17b} and the other is
MRI (magnetic resonance imaging) brain image (Fig. \ref{Fig-test-image} right) in medical imaging \cite{BrainWeb} -- for simulations. 
Credible regions and intervals are reported at $\alpha = 0.01$, i.e., 99\% Bayesian confidence.
We remark that the test M31 and MRI brain images used here, due to the limited space of the article, are just a showcase of the proposed UQ strategies for general 
inverse problems. Tests can certainly be performed analogously for other inverse problem applications with different types of images.

The parameter $\mu$ in \eqref{eqn:un-af} is estimated via algorithm \eqref{eqn:para-sel} with 10 iterations and
$\gamma, \beta, k = 1$. The MAP estimator at each iteration is computed via the forward-backward splitting used in \cite{CPM17b}. 
The automatically estimated parameter $\mu$ associated to the orthonormal basis and SARA dictionary for the analysis prior are reported in Table \ref{tab:snr-u},
which are then used for the synthesis prior. 

Fig. \ref{Fig-HPD} shows the computed threshold $\gamma^{\prime}_{\alpha}$ of the HPD credible regions, 
which tells us that the difference of the results between 
synthesis and analysis priors equipped orthonormal basis is negligible, but not for the SARA dictionary. 
This is consistent with the fact that over-complete basis can lead to a big different between the analysis and synthesis priors for
the original inverse problems. Moreover, we found from Table \ref{tab:snr-u} that the SNR of the point estimators (e.g. see Fig. \ref{Fig-lc-m31} (a)) 
using SARA library and the orthonormal basis are different either for the analysis or  the synthesis prior, which
again shows the difference due to the over-complete basis. 
Note that the prior with the over-complete SARA dictionary takes greater computation time than that with the orthonormal basis DB8 
since SARA contains more sub-bases.  

Fig. \ref{Fig-lc-m31} gives the results of the local credible intervals with respect to grid sizes of $10\times 10$  and $15\times 15$,
which shows that all the results are reasonable, and, more importantly, the differences between analysis and synthesis priors 
and between orthonormal basis and SARA dictionary are subtle (note that the results regarding the orthonormal basis are withdrew here due to the limited space and 
the negligible different in visual comparison). 
Fig. \ref{Fig-error} shows the error of the length of the local credible interval computed by MAP estimation compared to Px-MALA
(a state-of-the-art MCMC method \cite{M15}) corresponding to the analysis prior, with both orthonormal basis and SARA dictionary. 
From Fig. \ref{Fig-error}, we again see that both bases give very similar local credible interval error, using Px-MALA as the benchmark. 
Moreover, we see that the error of the MAP estimation is decreasing monotonically 
regarding different grid scales; particularly, lower than $\sim\!5\%$ when the grid scale is larger than $10\times 10$ (with ${\cal O} (10^5)$ orders of magnitude faster that
 Px-MALA). 

From the above results we conclude that selecting the regularisation parameter automatically can improve the effectiveness of  UQ 
strategies, applying more complex dictionaries improves the quality of point estimators, and the UQ results are consistent for 
different prior types and dictionaries. 
Future works will analyse the sensitivity (or robustness) of the UQ method with respect to the value of the regularisation parameter $\mu$.


\section{Conclusions}\label{sec:conclusions}
Analysing and quantifying uncertainties for inverse problems in image/signal processing is critical and very challenging
since the problems themselves are ill-posed and often high-dimensional.
In this article we presented a UQ methodology and investigated a series of UQ strategies based on MAP estimation for general image/signal processing problems. 
Particularly, in this article, two important new components -- automatic regularisation parameter selection and 
more general dictionaries/bases used in priors -- are considered to illustrate and improve the performance of these UQ strategies.
Moreover, the experimental results further strengthen the very promising performance of the proposed UQ strategies.
We emphasise again that these techniques are based on MAP estimation, and therefore can scale to
high-dimensional problems and problems with non-smooth objective functionals (e.g. sparsity-promoting $\ell_1$ prior).


\bibliographystyle{IEEEtran}
\bibliography{refs_xhcai}

\end{document}